\documentclass[%
reprint,
superscriptaddress,
 amsmath,amssymb,
 aps,
pra,
]{revtex4-1}
\usepackage{graphicx}% Include figure files
\usepackage{dcolumn}% Align table columns on decimal point
\usepackage{bm}% bold math
\usepackage{color}
\begin{document}

\title{Spectroscopy of the 1001 nm transition in atomic dysprosium}

\author{N. Petersen}
\affiliation{%
 QUANTUM, Institut f\"ur Physik, Johannes Gutenberg-Universit\"at, 55099 Mainz, Germany
}
\affiliation{%
Graduate School Materials Science in Mainz, Staudingerweg 9, 55128 Mainz, Germany}

\author{M. Tr\"umper}
\affiliation{%
 QUANTUM, Institut f\"ur Physik, Johannes Gutenberg-Universit\"at, 55099 Mainz, Germany
}

\author{P. Windpassinger}
\affiliation{%
 QUANTUM, Institut f\"ur Physik, Johannes Gutenberg-Universit\"at, 55099 Mainz, Germany
}
\affiliation{%
Graduate School Materials Science in Mainz, Staudingerweg 9, 55128 Mainz, Germany} 

\begin{abstract}
We report on spectroscopy of cold dysprosium atoms on the $1001\,\mathrm{nm}$ transition and present measurements of the excited
state lifetime which is at least $87.2(6.7)\,\mathrm{ms}$ long. Due to the long excited state lifetime we are able to measure the ratio of the excited state polarizability to the ground state polarizability at $1064\,\mathrm{nm}$ to be $0.828(0.129)$ by parametric heating in an optical dipole trap. In addition we measure the isotope shifts of the three most abundant bosonic isotopes of dysprosium on the $1001\,\mathrm{nm}$ transition with an accuracy better than $30\,\mathrm{kHz}$.

\end{abstract}

%\pacs{Valid PACS appear here}

\maketitle
\begin{figure}[ht!]
	\centering
		\includegraphics{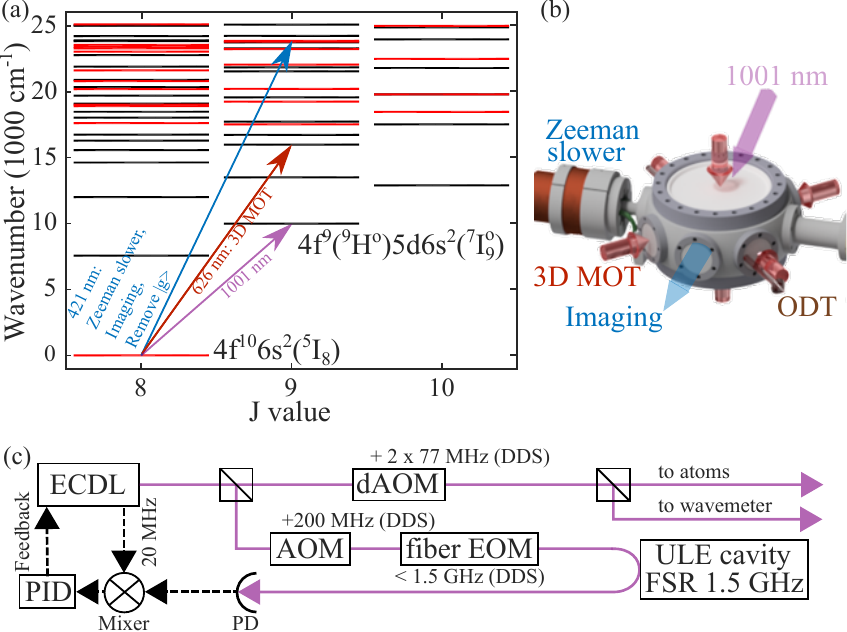}
		\caption{(a) Excerpt of Dy energy levels for total angular momenta $\mathrm{J}=8,9, \mathrm{and } \, 10$ \cite{NIST_ASD}. Even (odd) parity levels are drawn in red (black). The electron configuration of the ground state and of the excited state of the $1001\,\mathrm{nm}$ transition are written next to the corresponding levels. (b) Experimental setup. (c) $1001\,\mathrm{nm}$ laser system. DDS: direct digital synthesis, FSR: free spectral range.}
	\label{fig:figure1}
\end{figure}

Quantum gases of magnetic atoms enable the study of many-body physics with long-range, anisotropic interactions \cite{lahaye2009physics}. Due to their large magnetic moments, especially erbium (Er) and dysprosium (Dy) experiments became more prominent in recent years. This led to the realization of the extended Bose-Hubbard Hamiltonian \cite{baier2016extended}, the observation of the roton mode in the excitation spectrum of a dipolar Bose Einstein condensate \cite{chomaz2018observation}, and the discovery of self-bound quantum droplets \cite{ferrier2016observation,schmitt2016self}.
In addition to its large ground state magnetic moment, Dy features seven stable isotopes of which four have natural abundancies around 20\%. Due to its submerged and not completely filled 4f-electron shell, its energy spectrum, which is partially depicted in Fig.~\ref{fig:figure1}(a), is complex. Among the many possible transitions, there are at least two candidates for ultra-narrow-linewidth ground state transitions. One at $1322\,\mathrm{nm}$ with a predicted lifetime of $6.9\,\mathrm{ms}$ and one at $1001\,\mathrm{nm}$ with a predicted lifetime of $3\,\mathrm{ms}$  \cite{dzuba2010theoretical}. In this work we investigate the latter transition. 

Having an ultra-narrow-linewidth transition at hand enriches Dy quantum gas experiments with a versatile tool. Such transitions serve as sensitive probes for interactions between the atoms for example on lattice sites of an optical lattice \cite{scazza2014observation} or as probes for the external trapping potential e.g. to selectively probe different lattice sites in an optical superlattice.  Furthermore, because of its long lifetime, the excited state can be used as a second species in the experiment, whose population can be precisely controlled, e.g. to study Kondo-lattice physics \cite{foss2010probing,riegger2018localized}.
Ultra-narrow-linewidth transitions can also be used to probe the inner atomic potentials with a high sensitivity to investigate more fundamental physical questions. It has been proposed to use precise isotope shift measurements of two ultra-narrow-linewidth transitions of the same element to search for high-energy physics contributions to the inner potentials and for physics beyond the standard model \cite{delaunay2017probing,mikami2017probing}. In both cases the contributions to the inner potential could reveal themselves as a non-linearity in a King plot analysis \cite{king1963comments} of the two transitions. For this purpose the element needs to have at least four (zero nuclear spin) isotopes, which is fulfilled by Dy. Recently, the $1001\,\mathrm{nm}$ transition has for the first time been studied by laser spectroscopy and isotope shifts of all seven stable isotopes had been measured on the $40\,\mathrm{MHz}$ level \cite{studer2018laser}.

The experimental results presented here show that the lifetime of the excited state of the $1001\,\mathrm{nm}$ transition with $87.2(6.7)\,\mathrm{ms}$ exceeds the previous theoretical prediction \cite{dzuba2010theoretical}. In addition we refine the precision of the isotope shift measurements to the $30\,\mathrm{kHz}$ level for the three most abundant bosonic isotopes and present measurements of the excited state polarizability at the wavelength of $1064\,\mathrm{nm}$, which is commonly used for optical dipole trapping.

\section{EXPERIMENTAL SCHEME}

The experimental setup is depicted in Fig.~\ref{fig:figure1}(b).
A magneto-optical trap (MOT) operated on the $\Gamma_{626}=2 \pi \times 136\,\mathrm{kHz}$ transition from the ground-state at $626\,\mathrm{nm}$ is loaded from a Zeeman slower, which is using the $\Gamma_{421}=2 \pi \times 32\,\mathrm{MHz}$ broad $421\,\mathrm{nm}$ transition. Details of the experimental setup can be found in \cite{muhlbauer2018systematic}. Typically, $3\cdot 10^6$ $^{162}\text{Dy}$ or $^{164}\text{Dy}$ atoms or $3\cdot 10^5$ $^{160}\text{Dy}$ atoms are trapped and cooled to temperatures on the order of $30\,\mu\mathrm{K}$. The atoms can be transferred to a single beam optical dipole trap (ODT) at $1064\,\mathrm{nm}$. Absorption imaging is done in the horizontal plane in $45^\circ$ to the ODT beam using the $421\,\mathrm{nm}$ transition. The spectroscopy light at $1001\,\mathrm{nm}$ is generated by an extended cavity diode laser (ECDL), which is stabilized by the Pound-Drever-Hall method to a cylindrical cavity made of ultra-low expansion glass (ULE). We use a fiber EOM and an offset sideband locking technique to shift the laser frequency relative to the cavity resonances, which are spaced by $1.4961935(10)\,\mathrm{GHz}$. The ULE cavity is temperature stabilized to the zero-crossing temperature of the coefficient of thermal expansion and we achieve laser linewidths below $1\,\mathrm{kHz}$, and observe drifts of the stabilized laser frequency below $15\,\mathrm{kHz}/\mathrm{h}$. The $1001\,\mathrm{nm}$ light is scanned in frequency by a double pass acousto-optical modulator (dAOM). The radio frequencies driving the single pass and double pass AOMs and the fiber EOM are generated by direct digital synthesis. The circularly polarized $1001\,\mathrm{nm}$ spectroscopy beam is coming from top of the main chamber under a small angle to the z-axis. It is elliptically shaped and has diameters of $9.2\,\mathrm{mm}$ and $2.9\,\mathrm{mm}$ at the position of the atoms, where the larger half axis of the ellipse points along the ODT beam. From the same direction a resonant $421\,\mathrm{nm}$ beam can be applied to the atoms in the ODT to remove all ground state atoms from the trap. The magnetic field at the position of the MOT and the ODT is compensated during the application of the spectroscopy pulses.

\begin{figure*}[ht]
	\centering
		\includegraphics{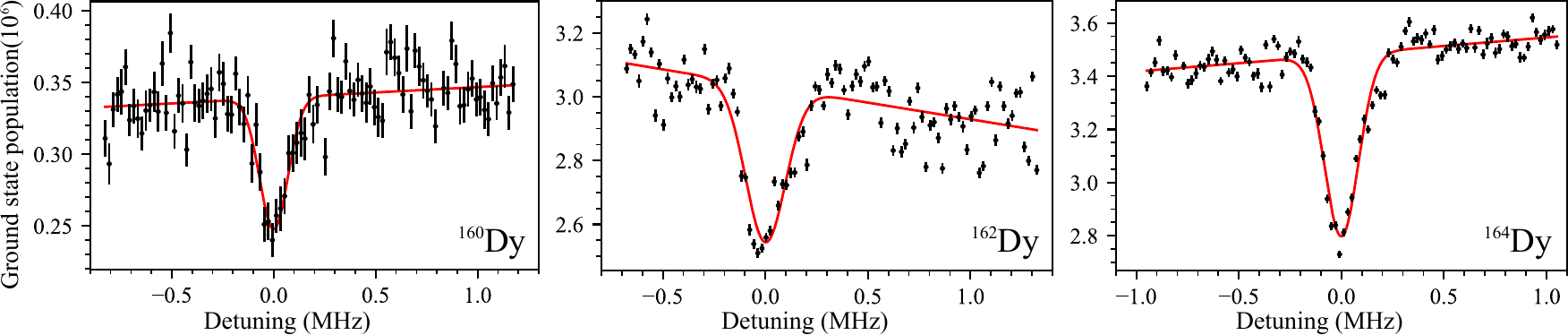}
		\caption{Exemplary spectra of the $1001\,\mathrm{nm}$ transition for $^{160}\mathrm{Dy}$, $^{162}\mathrm{Dy}$, and $^{164}\mathrm{Dy}$.}
	\label{fig:figure2}
\end{figure*}
 
\section{Measurement of the isotope shifts and absolute transition wavelength}
\label{sec:Exp results}

The spectroscopic measurements to obtain the isotope shifts are carried out in a pulsed manner. Atoms are released from the MOT and after $6\,\mathrm{ms}$ time of flight (TOF) a pulse of $6\,\mathrm{mW}/\mathrm{cm}^2$ light at $1001\,\mathrm{nm}$ is applied for $12\,\mathrm{ms}$ with a fixed detuning to the atomic resonance. After the pulse the remaining ground state population is measured by absorption imaging and then a new detuning is set and the measurement is repeated. This way the atomic resonance appears as a dip in the ground state population like in the three exemplary spectra presented in Fig.~\ref{fig:figure2}. The data points in each spectrum are obtained in sequence. Since there were slow drifts of the overall atom number the background of the spectra exhibits a slope. To determine the atomic resonance frequency from each spectrum an inverted Gaussian function with a linear offset is fitted to the data points. We reference the atomic resonances of $^{160}\mathrm{Dy}$ and $^{162}\mathrm{Dy}$ to the resonance of $^{164}\mathrm{Dy}$ and conduct a $^{164}\mathrm{Dy}$ measurement immediately before and after each measurement of one of the other isotopes. This way we can detect and account for drifts of the ULE cavity, which is the optical frequency reference in our setup. As result we obtain the following isotope shifts for $^{160}\mathrm{Dy}$ in Eq.~(\ref{eq:1}) and $^{162}\mathrm{Dy}$ in Eq.~(\ref{eq:2}) relative to $^{164}\mathrm{Dy}$:
\begin{equation} \label{eq:1}
\delta \nu_{160-164}= -2514.277(29)\,\mathrm{MHz,}
\end{equation}
\begin{equation} \label{eq:2}
\delta \nu_{162-164}= -1195.773(20)\,\mathrm{MHz.}
\end{equation}
The error budget is summarized in Table~\ref{tab:error budget} and takes contributions from ULE cavity FSR uncertainties, ULE frequency drifts, RF measurement uncertainties and fit errors into account. By using a wavelength meter (High Finesse WSU-30), we determine the absolute frequency of the transition for $^{162}\mathrm{Dy}$ to be:
\begin{equation} \label{eq:4}
\nu_{162}= 299.521643(30)\,\mathrm{THz}
\end{equation}
corresponding to a wavenumber of:
\begin{equation} \label{eq:3}
\tilde{\nu}_{162}= 9990.9666(10)\,\mathrm{cm}^{-1}\mathrm{.}
\end{equation}
The accuracy of the rubidium calibrated wavelength meter of $30\,\mathrm{MHz}$ is the dominant contribution to the measurement uncertainty by three orders of magnitude.
\begin{table}
\centering
\begin{tabular}{ l  c  c}
\textbf{Error budget} & {\boldmath$\delta \nu_{160-164}$} & {\boldmath$\delta \nu_{162-164}$} \\
\textbf{contribution} & {\boldmath$\mathrm{(kHz)}$} & {\boldmath$\mathrm{(kHz)}$} \\
\hline
ULE cavity FSR & 20 & 10 \\
ULE frequency drift & 14 & 6.7 \\
RF measurement uncertainty & 10 & 10 \\
Fit errors & 7.4 & 7.8 \\
\hline
\textbf{Total:} & \textbf{29} & \textbf{20} \\
\end{tabular}
 \caption{Summary of error contributions taken into account for the calculation of the total uncertainty of the measured isotope shifts. RF: radio frequency.}
 \label{tab:error budget}
\end{table}

\section{Measurement of the excited state lifetime}

\begin{figure}[ht!]
	\centering
		\includegraphics{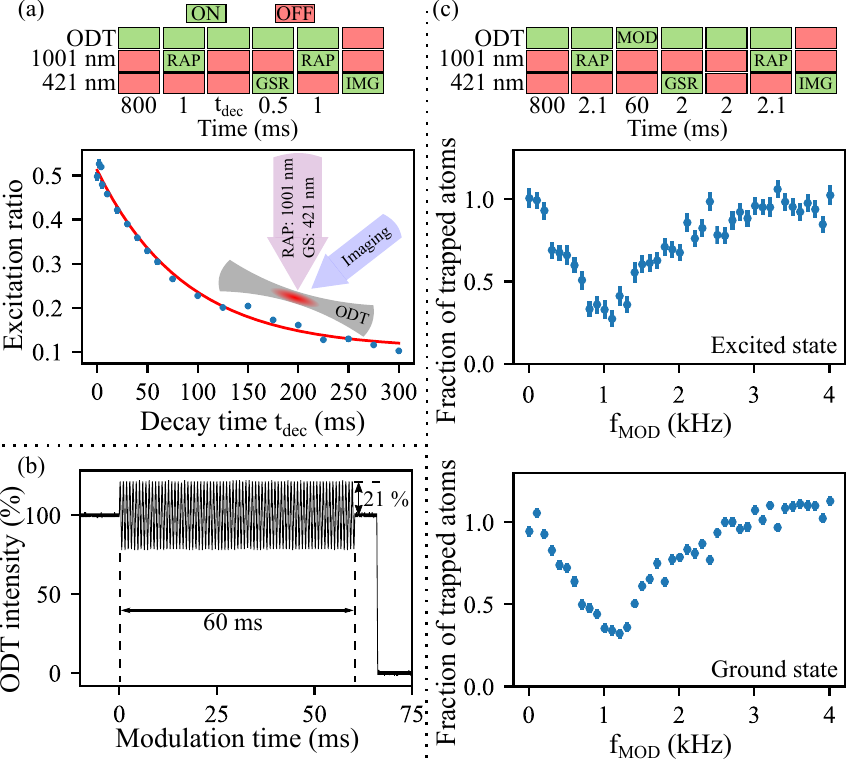}
		\caption{(a) Measurement sequence for the number of excited state atoms $N_e$ after a variable holding time. Where GSR refers to the ground state removing resonant $421\,\mathrm{nm}$ beam, MOD to modulation of the ODT intensity, and IMG to absorption imaging. The blue data points are the measured averaged excitation ratios $N_e/N_t$ versus the holding time after the first RAP in the sequence. The red line is the result of a fit with an exponential decay function to the data. The lifetime is $\tau_\mathrm{1001} \geq 87.2(6.7)\,\mathrm{ms}$. Inlet: Beam setup for the lifetime measurements. (b) Typical ODT intensity modulation at a frequency of $f_\mathrm{MOD}=1210\,\mathrm{Hz}$. (c) Sequence to measure $N_e$ after $60\,\mathrm{ms}$ modulation (MOD) of the ODT intensity. Parametric heating spectra for the excited state (top) and ground state (bottom).}
	\label{fig:figure3}
\end{figure}

The lifetime of the excited state is measured by the following sequence, which is outlined in Fig.~\ref{fig:figure3}(a). About $2.1 \cdot 10^5$ $^{162}\mathrm{Dy}$ atoms are trapped in the ODT after a holding time of $800\,\mathrm{ms}$. By applying a linear ramp of the laser detuning from $-300\,\mathrm{kHz}$ below the atomic resonance to $300\,\mathrm{kHz}$ above the atomic resonance in $984\,\mu\mathrm{s}$ with $4.88\,\mathrm{Hz}$ wide frequency steps and with a peak intensity of $450\,\mathrm{mW}/\mathrm{cm}^2$ about $71\,\%$ of the ground state population is transferred to the excited state of the $1001\,\mathrm{nm}$ transition by means of a rapid adiabatic passage (RAP). The mixture of ground and excited state atoms is trapped for a variable amount of time between $0.01\,\mathrm{ms}$ and $300\,\mathrm{ms}$ during which excited state atoms can decay back to the ground state. After this variable holding time, resonant $421\,\mathrm{nm}$ light is applied to remove the ground state population. Then another RAP transfers $71\,\%$ of the excited state atoms back to the ground state and subsequently the ground state atom number $N_e$ is measured by absorption imaging which is proportional to the number of excited state atoms before the second RAP pulse was applied. This sequence is repeated without applying the RAP pulses and the ground state removing $421\,\mathrm{nm}$ beam to obtain the total atom number $N_t$. For each holding time $N_e$ and $N_t$ are measured 15 times in turns and the resulting excitation ratios $N_e/N_t$ are averaged. The decay of the excitation ratio is plotted in Fig.~\ref{fig:figure3}(a). The order in which the excitation ratio was measured for different holding times was randomized. By fitting an exponential decay function to the data we find the lifetime of the excited state to be
\begin{equation} \label{eq:5}
\tau_\mathrm{1001} \geq 87.2(6.7)\,\mathrm{ms.}
\end{equation}
This is a lower limit for the lifetime since other effects that could decrease the excited state population over time like larger trap losses for the excited state atoms compared to the ground state atoms can not be fully excluded. Compared to the theoretical prediction of $\tau_\mathrm{1001,th.}=3\,\mathrm{ms}$ this is almost a factor 30 longer than expected \cite{dzuba2010theoretical}. 
\section{Measurement of the excited state polarizability}

Since the excited state lifetime is more than one order of magnitude larger than expected we are able to use parametric heating \cite{gehm1998dynamics,jauregui2001nonperturbative,friebel1998co} of excited state atoms in the ODT to measure the ratio of the excited state dynamic polarizability $\alpha_\mathrm{e}$ to the ground state dynamic polarizability $\alpha_\mathrm{g}$ at $1064\,\mathrm{nm}$.  For this purpose the intensity of the ODT beam is modulated for $60\,\mathrm{ms}$ with an amplitude of $21\,\%$ (Fig.~\ref{fig:figure3}(b)) and modulation frequencies $f_\mathrm{mod}$ ranging from $10\,\mathrm{Hz}$ to $4010\,\mathrm{Hz}$. The horizontal (vertical) beam radius is $w_h=28.3(5.9)\,\mu\mathrm{m}$ ($w_v=58.3(4.6)\,\mu\mathrm{m}$) at the position of the atoms and the beam power is approximately $11\,\mathrm{W}$. In the case of the excited state the sequence depicted in Fig.~\ref{fig:figure3}(c) is applied, where the ODT intensity modulation is switched on after a RAP transfers about $62\%$ of the atoms to the excited state. Then the ground state atoms are removed by resonant $421\,\mathrm{nm}$ light before a second RAP transfers part of the excited state population to the ground state and an absorption image is taken. The parametric heating spectrum for excited state atoms is depicted in Fig.~\ref{fig:figure3}(c) on the top and features a resonance at $f_e=1.043(60)\,\mathrm{kHz}$. Due to depletion of our atomic beam source the total ground state population is reduced to about $20 \cdot 10^3$ atoms in the ODT. 
The parametric heating resonance for ground state atoms under the same trapping conditions is measured by using the same sequence as for the excited state but without applying the RAPs and the resonant $421\,\mathrm{nm}$ light. The resulting spectrum is depicted in the lower half of Fig.~\ref{fig:figure3}(c) and it shows a resonance at $f_g=1.146(60)\,\mathrm{kHz}$.
From the resonances we obtain the ratio of the polarizabilities analog to \cite{ravensbergen2018accurate}:
\begin{equation}
\alpha_\mathrm{e}/\alpha_\mathrm{g}=(f_e/f_g)^2=0.828(0.129).
\label{eq: R}
\end{equation}
Theoretical calculations lead to $\alpha_\mathrm{e}/\alpha_\mathrm{g}=157\,\mathrm{a.u.}/181\,\mathrm{a.u.}=0.870$ \cite{dzuba2011dynamic}, while for the ground state 
\begin{equation}
\alpha_\mathrm{g}=184.4(2.4)\,\mathrm{a.u.}
\label{eq: alpha_g}
\end{equation}
is the experimentally determined value \cite{ravensbergen2018accurate}. From Eq.~(\ref{eq: R}) and Eq.~(\ref{eq: alpha_g}) we then obtain
\begin{equation}
\alpha_\mathrm{e}=153(24)\,\mathrm{a.u.}
\label{eq: alpha_e}
\end{equation}
for the excited state polarizability. The above results are in a good approximation independent of the exact trapping beam parameters if the ground and excited state populations are having similar density distributions in the trap and thus experience similar anharmonicity and beam aberrations. Obtaining precise values for the beam radii at the position of the atoms is difficult and usually prone to large relative errors. The above stated values for $w_h$ and $w_v$ were obtained from standard beam analysis with an attenuated ODT beam on a CCD camera and can have larger systematic errors than stated above. Nevertheless, we perform a measurement of the $1001\,\mathrm{nm}$ line shift for varying ODT beam powers and calculate the difference in polarizabilities to be:
\begin{equation}
\Delta \alpha=\alpha_\mathrm{e}-\alpha_\mathrm{g}=\frac{hc}{4 a_0^3} w_h w_v m_s = -21.5(5.0) \,\mathrm{a.u.},
\end{equation}
where $h$ is the Planck constant, $c$ is the speed of light, $a_0$ is the Bohr radius and $m_s=-38.9(2.4)\,\mathrm{kHz/W}$ is the observed slope of the line shift per beam power. A reduction of the mean intensity experienced by the atoms due to their spread around the potential minimum is taken into account in form of a correction factor in the calculation of $m_s$. With Eq.~(\ref{eq: alpha_g}) we obtain $\alpha_\mathrm{e}=162.9(5.5) \,\mathrm{a.u.}$. Since the vector and tensor polarizabilities of both states are expected to be two orders of magnitude smaller than the scalar polarizabilities, dependencies on ODT beam polarization and the atomic azimuthal quantum number were neglected in the considerations \cite{PolDytheory,ravensbergen2018accurate}.
\section{CONCLUSIONS}

We have measured the relative isotope shifts of the three most abundant bosonic isotopes of dysprosium on the $1001\,\mathrm{nm}$ transition with  an accuracy better than $30\,\mathrm{kHz}$ while the absolute frequencies were determined with an uncertainty of $30\,\mathrm{MHz}$. In addition, we have determined a lower boundary for the excited state lifetime which is more than one order of magnitude larger than expected from theoretical predictions \cite{dzuba2010theoretical}. The dynamical polarizabilitiy of the excited state was determined relatively to the ground state dynamical polarizabilitiy and the ratio is in fair agreement with theory \cite{dzuba2011dynamic}.

\section{ACKNOWLEDGEMENTS}

The authors would like to thank Florian M\"uhlbauer, Lena Maske, Gunther T\"urk and Carina Baumg\"artner for their contributions to the experiment and the group of Klaus Wendt for their support, advice and joint use of their wavelength meter. We thank Dmitry Budker for his very appreciated advice during the initial search for the $1001\,\mathrm{nm}$ transition.
We gratefully acknowledge financial support by the DFG-Grossger\"at INST 247/818-1 FUGG, and the Graduate School of Excellence MAINZ (GSC 266).

\bibliographystyle{unsrtnat}
\bibliography{1001_v2}{}

\end{document}